# Imaging and controlling electron motion and chemical structural dynamics of biological system in real time and space

Ligong Zhao[1,2,†], Mohamed Sennary[1,†], Dina Hussein[1], Anaísa Coelho[3], Tingting Yang[3] Mohamed Y El-Naggar [3,4,5], and Mohammed Th. Hassan[1]*

[1]Department of Physics, University of Arizona, Tucson, AZ 85721, USA.
[2] Center for Attosecond Science and Technology, Xi'an Institute of Optics and Precision Mechanics, Chinese Academy of Sciences, Xi'an 710119, China.
[3]Department of Physics and Astronomy, University of Southern California, Los Angeles, CA 90089, USA.
[4]Department of Chemistry, University of Southern California, Los Angeles, CA 90089, USA
[5]Department of Biological Sciences, University of Southern California, Los Angeles, CA 90089, USA

* Corresponding authors: mohammedhassan@arizona.edu
†These authors contributed equally to this work.

**Ultrafast electron microscopy (UEM) has found widespread applications in physics, chemistry, and materials science, enabling real-space imaging of dynamics on ultrafast timescales. Recent advances have pushed the temporal resolution of UEM into the attosecond regime, giving rise to attomicroscopy—a technique capable of directly visualizing electron motion. In this work, we extend the capabilities of this powerful imaging tool to investigate ultrafast electron dynamics in a biological system by imaging and controlling light-induced electronic and chemical changes in the conductive network of multicellular cable bacteria. Using electron energy loss spectroscopy (EELS), we first observed a laser-induced increase in π-electron density, accompanied by spectral peak broadening and a blueshift—features indicative of enhanced conductivity and structural modification. We also traced the effect of ultrafast laser pumping on bulk plasmon electron oscillations by monitoring changes in the plasmon-like resonance peak. Additionally, we visualized laser-induced chemical structural changes in cable bacteria in real space. The imaging results revealed carbon enrichment alongside a depletion of nitrogen and oxygen, highlighting the controllability of chemical dynamics. Moreover, time-resolved EELS measurements further showed a picosecond-scale decay and recovery of both π-electron and plasmonic features, attributed to electron–phonon coupling. In addition to shedding light on the mechanism of electron motion in cable bacteria, these findings demonstrate ultrafast modulation and switching of conductivity, underscoring their potential as bio-optoelectronic components operating on ultrafast timescales.**

**Introduction**

The advancement of Ultrafast Electron Microscopy (UEM) has enabled the real-time observation of laser-induced ultrafast atomic and molecular dynamics (*1-8*). More recently, enhancements in temporal resolution—reaching the extreme attosecond time scale—and the development of attomicroscopy, have made it possible to directly trace electron motion (*9-11*). A key strength of UEM and attomicroscopy lies in their ability to bridge ultrafast dynamics with the intrinsic properties of the sample under investigation, offering new opportunities to design and develop ultrafast and petahertz optoelectronic technologies (*12-23*).

While UEM has been widely applied to image ultrafast dynamics in materials science, physics, and chemistry (*7, 24-34*), its applications in biology have been largely limited to the studies of mechanical movement and surface structure (*35-37*). For example, UEM has been used to detect picometer-scale movements in thin films of photoresponsive insulin amyloid fibrils embedded in vitreous ice on nanosecond timescales (*35*). It has also enabled the noninvasive visualization of oscillatory dynamics in freestanding amyloid nanocrystals (*38, 39*), as well as investigations into the mechanical properties of freestanding DNA nanostructures (*40*). The development of Photo-Induced Near-Field Electron Microscopy (PINEM) (*6*) has further allowed direct imaging of various biological surfaces and membranes, including protein vesicles, whole *Escherichia coli,* and eukaryotic cells (*36, 41*).

In this study, we expand the capabilities of UEM and attomicroscopy to demonstrate, for the first time to our knowledge, the imaging of electron dynamics and light-induced chemical structure modifications in a biological system (cable bacteria).

Cable bacteria are filamentous microorganisms capable of long-distance electron transport and have drawn considerable interest in microbial ecology due to their unique ability to conduct electricity over centimeter-scale distances (*42-47*). These multicellular filaments, composed of thousands of end-to-end cells, gain energy from the long-distance electron conduction process by coupling sulfide oxidation in deeper sediment layers to oxygen reduction near the sediment-water interface(*45*). Fast charge transport over these distances was previously thought impossible in biological materials. However, an emerging body of evidence suggests that electron transport proceeds along a periplasmic network of conductive nanofibers that incorporate novel metalloprotein nickel cofactors (*48-50*). While long-distance electron transport through proteins is often described with a multi-step hopping process between redox cofactors, the remarkably high conductivity of cable bacteria potentially suggests that hopping events may be assisted by delocalization of electrons across islands of tightly stacked cofactors or conjugated molecules(*49*). It has also been speculated that heme cofactors of cytochromes may facilitate the transfer of electrons into and out of the conductive nanofibers (*48, 51*). These electron transport mechanisms differ fundamentally from the cytochrome-based extracellular electron conduits of other electroactive bacteria (*52-54*), representing a novel biological conduction system. As such, cable bacteria are promising candidates for the development of bioelectronic devices, with potential applications in medical technologies.

In this work, we study the conductive nanofiber network of cable bacteria as bio-nanowires, focusing on ultrafast laser-induced intrinsic electronic dynamics and their associated conductivity using electron energy loss spectroscopy (EELS). By analyzing the evolution of the π-electron and bulk plasmon-like peaks in the low-energy EELS spectra (*55*) under varying pump laser powers, we gained insights into the electron behavior and its connection to changes in chemical structure and electrical conductivity of the bacteria's conductive nanofibers. Spectral parameters—including peak area, position, and full width at half maximum (FWHM)—were used to quantitatively characterize these dynamics. Additionally, we employed elemental mapping to visualize real-space chemical modifications of the three primary elements in cable bacteria—carbon, nitrogen, and oxygen. By systematically varying the pump laser power, we demonstrated control over the chemical structure and correlated these changes with the observed electron dynamics. Furthermore, we examined ultrafast electron-phonon coupling and its impact on conductivity modulation in cable bacteria. These findings offer the first demonstration of controllable, light-induced electron and chemical dynamics in cable bacteria, opening new avenues for the development of ultrafast bio-optoelectronic devices with potential applications in biomedicine.

**Materials and Methods**

In our attomicroscopy setup (*9*), electrons are generated within the microscope either in continuous wave mode via thermionic heating of the photocathode source or as ultrafast electron pulses by illuminating the photocathode with ultrafast UV pulses for time-resolved EELS measurements. The present study focuses on ultrafast phonon–electron dynamics occurring on a picosecond time scale; therefore, the optical gating (*10, 11*) and attosecond resolution capabilities of the attomicroscope setup were not utilized. The electron beam is focused onto the sample using a magnetic lens, with the beam diameter set to 100 μm. The pump near-infrared (NIR) pulse is generated through nonlinear propagation of the output pulse from the OPCPA laser system. This NIR pulse, centered at 800 nm with an estimated duration of 20 fs, is directed and focused onto the cable bacteria sample inside the microscope using a focusing mirror. The beam diameter on the sample is approximately 200 μm, sufficient to cover the entire cable bacterium, which is several micrometers in length. The power of the pump beam is adjusted using a neutral density filter. The temporal delay between the laser pulse and the UV (and thus the electron) pulse is controlled via a delay stage placed in the laser beam path. Our microscope is equipped with an electron energy spectrometer for EELS spectrum acquisition.

Marine sediment containing cable bacteria was collected from the Upper Newport Bay Conservation Area. In the laboratory, the sediment was homogenized, sieved, autoclaved, and carefully transferred into glass jars, ensuring that no air bubbles were trapped. After cooling, cable bacteria were obtained by placing a piece of non-autoclaved sediment under a dissection microscope and extracting individual filaments using homemade glass hooks. The bacteria were then inoculated into sterilized sediment. The jars were incubated in a water bath at 15 °C, fully submerged in circulating, air-saturated 2.6% seawater. After 5 weeks of incubation, small sediment cores were collected from the incubation jars using plastic straws (8 mm diameter, 80 mm length) and transferred to small petri dishes containing 2.6% seawater. Long filaments were hand-picked under a dissection microscope using glass hooks made from capillary tubes and washed at least three times in Milli-Q water droplets to remove sand and debris. All extractions were performed inside a glove bag under a nitrogen atmosphere to maintain anoxic conditions. Bundles of filaments were thoroughly washed and deposited in small droplets of Milli-Q water. Then, the samples were sealed inside glass jars to maintain an anoxic environment and transferred into an anaerobic chamber under the nitrogen atmosphere. All subsequent steps were carried out entirely inside the anaerobic chamber under nitrogen to preserve anoxic conditions. This includes the deposition of droplets containing cable bacteria onto a PELCO® ultrathin carbon film supported by a lacey carbon film on a 300-mesh gold grid, the addition of 4 μL of 1% (w/v) SDS to completely cover the grid surface. The SDS-treated grids were then placed in a humid chamber and incubated for 6 hours. After SDS incubation, the samples were washed three times with Milli-Q water and allowed to dry, under strictly anoxic conditions throughout. The extracted sheaths containing the nanofiber network of cable bacteria were subsequently sealed inside the chamber and were only opened when loaded to the microscope to perform the experiments.

**Results**

In our first experiment, we illuminated the extracted nanofiber network of cable bacteria with the ultrafast near-infrared (NIR) laser. The TEM image of the sample is shown in Fig. 1A. Then, we recorded the low-energy electron energy loss spectra (EELS) of the sample using a continuous-wave thermionic electron beam at various pump laser powers. The resulting spectra (ranging from 0 to 90 eV) are shown in Fig. 1B. These spectra reveal two distinct peaks at 3.6 eV and 23.1 eV.

The two features are consistent with previous EELS spectra of cellular biomolecules (e.g. protein, DNA) due to collective excitations of valence electrons, namely a broad ~23 eV plasmon-like peak and another peak (< 10 eV) associated with the excitation of π-states (*56-59*). The first peak at 3.6 eV therefore indicates the presence of π-electrons such as those occurring in peptide bonds, aromatic side groups, cofactors, or other conjugated structures associated with the conductive nanofiber network. These π-electrons represent collective excitations and may contribute to electrical conduction if they are sufficiently delocalized. The integration of this peak (*A*, area under the curve) increases exponentially with rising pump laser power up to 30 mW, after which it saturates (see Fig. 1C). This nonlinear behavior suggests an increase in the number of excited π-electrons as the pump power increases, up to a threshold, due to multiphoton excitation of NIR photons. Additionally, the peak broadens with increasing power—its full width at half maximum (FWHM) increases from 3.9 eV at 0 mW to 5.1 eV at 50 mW (as shown in Fig. 1D)—indicating reduced coherence and shorter lifetimes of the excited π-electrons. This broadening might be due to enhanced electron-phonon interactions, driven by laser-induced atomic vibrations and local thermal heating. Moreover, a gradual blueshift of the peak, from 3.6 eV to 4.2 eV, is also observed with increasing pump power (Fig. 1E). This blueshift is potentially due to light-induced modifications in the electronic structure of the system, such as the formation of new π-bonds.

In organic conductors, π-electrons often serve as the primary charge carriers. While the electron transport mechanism in cable bacteria is not yet known, their unusually high conductivity and low activation energy were previously invoked to suggest a novel mechanism where hopping between 'relay segments' (e.g. π-conjugated molecules or tightly stacked cofactors) is assisted by substantial electron delocalization within the relay segments (*49, 60*) Regardless of the precise identity and structure of the electron transfer sites, we propose that the π-electrons observed here may enhance the electron mobility and contribute significantly to overall conduction along the nanofibers. Assuming the conductivity $\sigma$ could be similar to conducting wires, it is given by

$$\sigma = \frac{ne^2 \tau}{m_e} \qquad (1)$$

Where $\tau = \frac{1}{\gamma}$ is the electron relaxation time, $(n)$ is the electron density, $(e)$ is the electron charge, $m_e$ is the effective mass of electron.

In EELS, area under the peak $(A)$ is proportionally related to the oscillator strength and the number of free electrons $(n)$ (*61*), and consequently the conductivity $\sigma$, so

$$\sigma \propto A \qquad (2)$$

The width (FWHM) of the peak ($E_{FWHM}$) is inversely proportional to the electron relaxation time $\tau$ and conductivity $\sigma$ (see eq. 3). The narrower the peak means a smaller $\gamma$, indicating less energy loss due to electron scattering and increase in the conductivity. Hence,

$$\sigma \propto 1/E_{FWHM} \tag{3}$$

Accordingly, the conductivity before illuminating the cable bacteria nanofibers with laser beam is

$$\sigma_0 \propto A_0/(E_{FWHM})_0 \tag{4}$$

The change in conductivity due to illumination of the nanofiber at a given laser power $p$ can be expressed as

$$\sigma_p \propto A_p/(E_{FWHM})_p \tag{5}$$

So, the relative change of the conductivity $\sigma_p/\sigma = \frac{A_0/A_p}{(E_{FWHM})_0/(E_{FWHM})_p} \tag{6}$

Accordingly, from the results shown in Fig.1C and D, we estimated laser-excitation of π-electrons would enhance the conductivity by ~15%.

The second peak at 23.1 eV represents the bulk plasmon-like electron oscillations (Fig. 1B), which is raised due to the oscillation of the electrons through the bacteria filament. These electron oscillations occur at a specific frequency, known as the plasma frequency, which is determined by the density of free electrons:

$$\omega_p = \sqrt{\frac{ne^2}{\varepsilon_0 m_e}} \tag{7}$$

Bulk plasmon electrons are not a separate class of electrons but are instead the same free electrons involved in collective, wave-like oscillations. When a bulk plasmon is excited—by an electromagnetic wave or an electron beam—the electrons oscillate coherently at the plasma frequency ($\omega_p$), and these dynamics can provide insight into the conductivity. However, because this motion is purely oscillatory, the average displacement of the electrons over time is zero, resulting in no net charge transport. Consequently, bulk plasmon electrons likely contribute minimally to long-distance electron transport in the nanofibers of cable bacteria. Our results show that the area under the curve ($A$) of the bulk plasmon-like peak at 23.1 eV decreases exponentially with increasing pump laser power, reaching approximately 22% of its original value at 50 mW (Fig. 1F). This reduction suggests a decrease in the number of bulk plasmon valence electrons density $n$, likely due to laser-induced structural modifications in the bacterial material. Interestingly, the coherence of the remaining plasmonic electrons appears to improve, as indicated by a ~13% decrease in the full width at half maximum (FWHM) of the peak (Fig. 1G). This narrowing implies enhanced electron mobility and improved conductivity within the conductive nanofibers of cable bacteria. Furthermore, the plasmon-like peak exhibits a noticeable redshift with higher pump laser powers, shifting from 23.1 eV to 22.2 eV, as shown in Fig. 1H. While a redshift is typically associated with decreasing the electron density (shown in Fig. 1F), chemical structural changes, or thermal expansion due to induced vibration and phonon dynamics in the system.

To explore the potential light-induced chemical structural dynamics underlying the π-electron and bulk plasmon valence electron behavior observed in Fig. 1, we conducted elemental mapping using electron energy loss spectroscopy (EELS) to image real-space changes in the primary chemical elements of the cable bacteria: carbon ($C$), nitrogen ($N$), and oxygen ($O$) atoms. Please note that the Fe, Ni, or S were not observed in our measurements since our microscope is more modified to ultrafast measurements. Hence, the signal to noise ratio at high energy EELS spectra is not good enough to resolve the peaks of these elements. The elemental mapping images and corresponding EELS spectra—obtained after background subtraction—are presented in Fig. 2. As shown in Fig. 2A, the contrast of carbon increases with rising pump laser power, indicating a possible enhancement in carbon-based bonding or structure. In contrast, nitrogen (Fig. 2B) and oxygen (Fig. 2C) signals gradually diminish at higher laser powers, suggesting a loss or redistribution of these elements. Notably, despite these compositional changes, the overall morphology of the sample remains intact in real-space imaging. This demonstrates our ability to induce controlled chemical modifications in the bacteria using light, without causing observable physical damage to the sample in real space.

First, we quantitatively analyzed the dynamics of the three elements by integrating the EELS spectra to estimate the total number of electrons associated with each element and plotted this as a function of pump laser power for carbon, nitrogen, and oxygen in Fig. 3A, B, and C, respectively. These curves reveal nonlinear trends: the electrons count increases for carbon and decreases for nitrogen, and oxygen, which likely related to the behavior observed in the π-electron excitations and bulk plasmon-like features. It is important to note that the carbon and nitrogen peaks are recorded on the same spectrum, hence the increase of the scattered electron at the carbon atom region is due to the reduction of the scattered number of electrons at the nitrogen atom spectral region, not due to increase in carbon atom quantity since the experiment is done under vacuum inside the microscope.

To further investigate light-induced chemical changes in real space, we analyzed the elemental mapping images for carbon, nitrogen, and oxygen (Fig. 3D, E, and F). We selected six regions of interest (ROIs) and tracked their intensity variation as a function of pump laser power (Fig. 3G, H, and I). ROIs 1 and 3 are located at the membrane, while ROIs 2, 4, 5, and 6 are positioned at the centre of the cable bacteria filaments. At no laser illumination (0 mW pump power), the data suggest that carbon and oxygen are more concentrated at the centre of the filaments, whereas nitrogen is more prominent in the membrane regions. The laser-induced changes in carbon (Fig. 3D and G) are relatively uniform across all ROIs, suggesting a homogeneous modification in electron density. In contrast, changes in nitrogen and oxygen distributions are more spatially heterogeneous. For nitrogen (Fig. 3E and H), the signal decreases more significantly in the central regions (ROIs 2, 4, 5, and 6) compared to the membrane (ROIs 1 and 3), whereas for oxygen (Fig. 3F and I), the opposite tendency is observed.

Second, to gain deeper insight into the chemical dynamics qualitatively, we analysed the EELS spectra of carbon, nitrogen, and oxygen at various pump laser powers, as shown in Fig. 4A, B, and C, respectively. The carbon EELS spectrum (Fig. 4A) displays three distinct peaks: $C_1$ at 287 eV, attributed to π-bond electrons; $C_2$ at 298 eV, corresponding to σ-bond electrons; and $C_3$ at 321 eV, associated with excited electrons from sp³-hybridized carbon atoms (62). Beyond a general increase in electron count across the spectrum, both the $C_1$ and $C_3$ peaks become more prominent with increasing the pump laser power. This suggests enhanced formation of π-bonds as the laser illumination modifies the chemical structure of the cable bacteria nanofibers. The strengthening of the $C_1$ peak is consistent with the observed increase in the π-electron peak intensity in the low-energy EELS spectra (Fig. 1C). Additionally, the $C_3$ peak exhibits a blueshift at higher pump powers, which may indicate changes in the configuration of newly formed π-bonds. The nitrogen EELS spectrum (Fig. 4B) contains two peaks—$N_1$ and $N_2$—representing π- and σ-bond contributions, respectively. The oxygen spectrum (Fig. 4C) features a single peak, $O_1$, corresponds to the C–O σ-bond (63). At high pump laser powers, these peaks in the nitrogen and oxygen spectra significantly diminish or nearly vanish, suggesting possible ionization or evaporation of N and O atoms from the cable bacteria nanofibers. This interpretation is supported by the elemental mapping results shown in Fig. 2B and C. The reduction in nitrogen and oxygen content may also contribute to the decrease in the bulk plasmon-like peak at 23.1 eV observed in Fig. 1F. Together, these results demonstrate the ability to manipulate the chemical structure of cable bacteria nanofibers using ultrafast near-infrared laser pulses, enabling controlled irreversible modification at the molecular level without apparent structural damage.

In the second experiment, we studied the ultrafast electron reversible dynamics using time-resolved low-loss EELS. Ultrafast electron pulses (triggered by UV laser on the cathode) probed the sample at varying time delays of the triggering NIR pump laser pulse (power is set at 30 mW). Then, we integrated the total number of electrons (A) for the π- (at 3.6eV) and plasmon-like (at 23.1 eV) electron peaks over time. Their dynamics are shown in Fig. 5A and B, respectively. Both dynamic curves are fitted using a biexponential function and the fitting curves are shown in black (Fig. 5A) and red (Fig. 5B) lines. As retrieved from the fitting results, the π-electron peak shows a rapid decay in A (drop until ~90%) on a ~1.9 ps time scale followed by the start the recovery within ~3 picoseconds (as shown in Fig. 5A). The full recovery to the ground state might takes few tens or hundreds of picoseconds, out of the scope of this study. The bulk plasmon-like peak integration dynamics, in Fig. 5B, show faster decay time on the order of ~850 fs and slower recovery in ~3.7 ps. This behavior likely arises from electron–phonon coupling and the molecular lattice vibrations disrupting coherent electron oscillations. These vibrations reduce carrier mobility by scattering electrons and altering tunneling barriers. This reduces conduction (almost ~10%), paralleling the behavior in organic semiconductors where such effects limit mobility(64-66). Thus, as shown in Fig. 5, cable bacteria nanofiber conductivity can be modulated in real time—switching from high to low states of conductivity (by 10% contrast) within 2 picoseconds (0.5 THz) or less—demonstrating the potential for ultrafast bio-optoelectronic switching and electronics with terahertz speed.

**Discussion**

Cable bacteria represent a unique biological system where cells gain energy by carrying electrons generated from sulfide oxidation in deeper sediment layers to drive oxygen reduction near the sediment-water interface. Achieving this unusual 'electric metabolism' requires chemical structures and mechanisms of electron motion that give rise to macro-scale long-distance electron transport along the parallel network of nanofibers residing in the cell envelope. While the precise identity of the protein charge carrier sites giving rise to this conductivity is not yet known, studies have investigated the possible involvement of Fe-containing hemes or sulfur-coordinated Ni cofactors reminiscent of Ni-bis(dithiolene) complexes (67, 68) However, a better understanding of the electron transport mechanism will benefit from *in situ* new techniques to examine the ultrafast electron and chemical dynamics, including of associated π-electrons, as demonstrated in this study.

Upon illumination of the cable bacteria's nanofiber network with an ultrafast near-infrared (NIR) laser pulse, the conductivity is enhanced due to an increase in π-electron density, which is probed using an electron beam in conjunction with bulk surface plasmon electrons. We monitored this change by recording electron energy loss spectroscopy (EELS) spectra in both low-energy and K-edge spectral regions.
The π-electron density can be regulated by varying the power of the NIR laser beam, which increases proportionally with laser intensity. Simultaneously, the number of excited plasmon electrons decreases (Fig. 1). This electronic dynamic behavior correlates with changes in the chemical structure of the cable bacteria nanofibers induced by the laser beam (Figs. 2, 3 and 4). At higher powers, oxygen and nitrogen atoms begin to ionize and evaporate, reducing their concentration within the bacteria. This phenomenon accounts for the decrease in valence electron oscillations, as evidenced by the diminishing bulk plasmon-like peak. As the C–O and C–N bonds break, carbon atoms form more π-bonds together, as indicated by the EELS spectra in Fig. 4A.

We employed elemental mapping imaging techniques to visualize changes in the nanofiber network under laser illumination, achieving sub-micrometer spatial resolution (Figs. 2 and 3). Despite these structural modifications, the morphology of the cable bacteria nanofibers remains preserved, suggesting that the chemical alterations do not result in physical sample damage.
The increased π-electron density enhances the electrical conductivity of the nanofibers, indicating potential for developing biophotonic devices.
While the first experiment investigates the permanent changes of the electron dynamics and chemical structure in nanofibers induced by the laser beam, the second experiment focuses on studying the real-time reversible dynamic changes in π-electron density and bulk surface plasmons that induced by the 30 mW laser beam. Our findings reveal that the π-electron density—and thus the conductivity of cable bacteria nanofibers— first decreases by 10% within 1.9 ps after the excitation, before recovering in 3 ps. Bulk plasmon-like electrons exhibit a faster decay but similar temporal recovery behavior. Laser pulse excitation induces atomic vibrations within the bacterial molecular structure, generating phonon waves on a picosecond timescale. This electron–phonon coupling opens new decay channels, reducing electron density and conductivity over the same period.
Consequently, the current—correlated with the conductivity dynamics—is expected to switch from a maximum (ON state, 100% at 0 ps delay) to a minimum (OFF state, 90% at 1.9 ps, if the threshold is set at this value), demonstrating the feasibility of utilizing the conductive components of cable bacteria for ultrafast bio-optoelectronic switches and transistors.
Our exploration of ultrafast phonon dynamics is crucial for optimizing the performance of these conductive nanofibers in bioelectronic applications, ensuring minimal energy loss. Moreover, understanding the phonon dynamics may enable the design of materials with tailored thermal conductivity, decisive for the stability of implantable sensors and environmental monitors. Furthermore, since phonon dynamics influence vibrational modes that govern mechanical stability and elasticity, analyzing these modes ensures the nanofibers can withstand physical stress in wearable or biodegradable devices. Also, in potential bioelectronic applications, where cable bacteria components interface to electrodes (e.g., in biosensing), phonon dynamics influence how electrical signals couple with vibrational modes, directly impacting signal transduction efficiency. Gaining insight into these processes ensures effective communication between biological and electronic components.

In this study, we investigated and demonstrated control over the electron dynamics and chemical structure changes induced by ultrafast laser pulses in cable bacteria nanofibers. Laser-induced electronic excitations were found to enhance the conductivity of the nanofibers. Additionally, laser illumination altered the chemical structure of key elements within the bacteria, which we tracked with sub-micrometer spatial resolution across the surface and membrane. We also examined ultrafast laser-induced electron–phonon coupling and its impact on the conductivity of the cable bacteria. These findings provide important insights, both towards understanding the mechanism(s) of electron transport in cable bacteria, and for harnessing this unique biological system in the development of ultrafast bio-optoelectronic devices.

## Acknowledgments

This project is funded by Gordon and Betty Moore Foundation Grant GBMF 7938 to M. Hassan. We are also grateful to the W.M. Keck Foundation for supporting this project with a Science and Engineering award given to M. Hassan. The El-Naggar group was supported by the Gordon and Betty Moore Foundation grant 10148 and W.M. Keck Foundation award 8626. The sediment samples containing cable bacteria were collected under the Scientific Collecting Permit S-222970004-22297-001, which was granted by the State of California – Department of Fish and Wildlife. We are also grateful to Ms. Magdalene MacLean (University of Southern California) for assistance with sediment collection and preparation of samples.

## Author contributions:

L.Z., and M.S. conducted the experiments and analyzed the data with help from D.H. D.H. T.Y and A.C. arrange and prepared the cable bacteria sample. M.Y.E and M.Th.H. conceived, supervised, and directed the study. All authors discussed the results and their interpretation and wrote the manuscript.

**Figures and Tables**

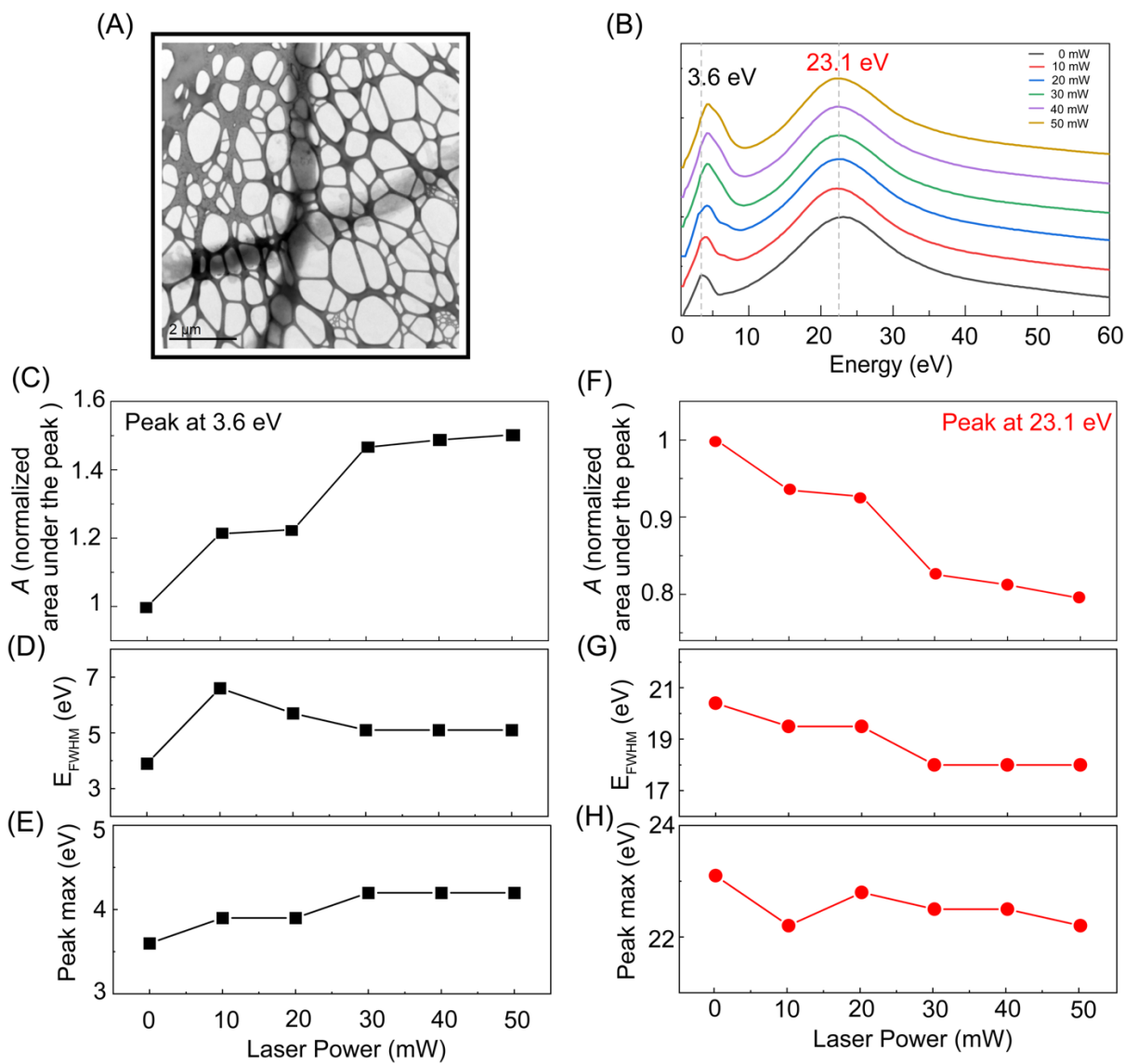

**Fig. 1. The electron dynamics in cable bacteria**. **(A)** TEM image of the extracted sheaths containing the nanofiber network of cable bacteria. **(B)** Low-energy EELS spectra of the nanofiber network at different pump power of the ultrafast NIR-laser pulse. **(C-E)**, and **(F-H)** are the area under the curve (*A*), the full-half width maximum ($E_{FWHM}$) and the max of the peak energy, of the π-electron excitations peak at 3.6 eV and bulk plasmon-like peak at 23.1 eV change as a function of the pump laser power, respectively.

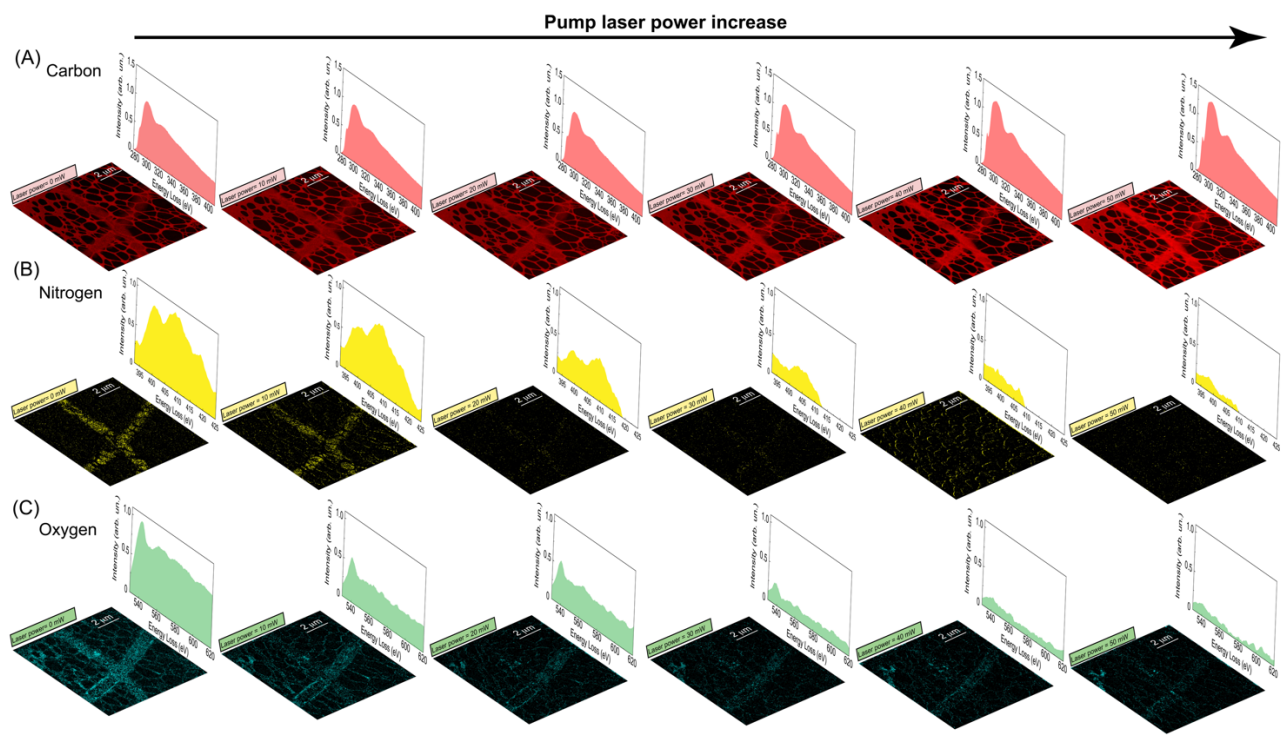

**Fig. 2. Imaging the chemical structure change of cable bacteria nanofiber network in real space. (A, B and C)** Elemental mapping and corresponding K-Edge EELS spectra of the carbon, nitrogen and oxygen atoms, respectively, at different pump laser power (from 0-50 mW).

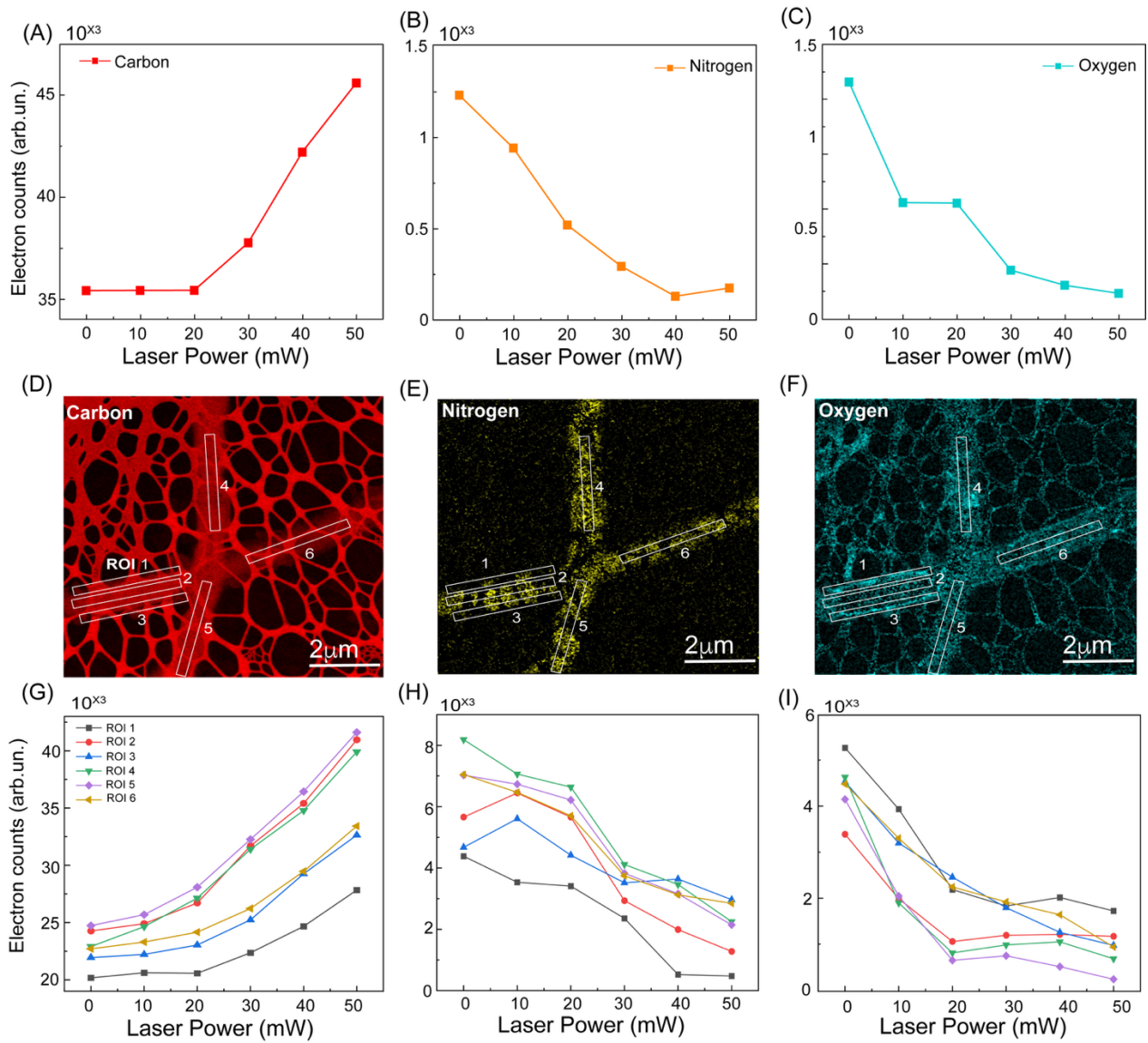

**Fig. 3. Sub-micrometer tracing of the chemical structure change in cable bacteria nanofiber network in real space. (A-C)** Number of electrons counts change of the carbon, nitrogen, and oxygen EELS peaks at different laser powers, respectively. (D,G), (E,H), and (F,I) tracing the intensity of the carbon, nitrogen and oxygen change at six different regions of interest (ROI) of the cable bacteria sample.

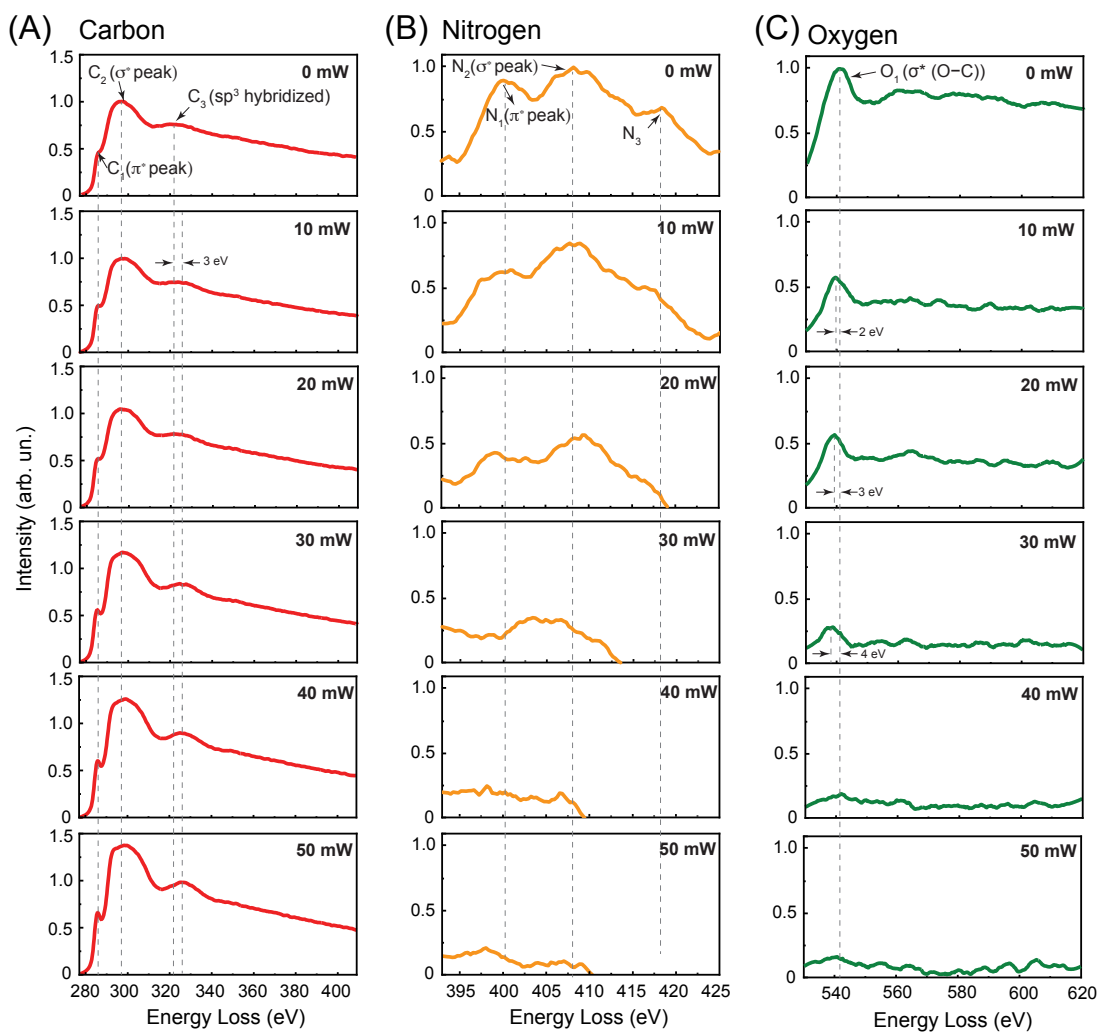

**Fig. 4. EELS spectra of the main chemical elements in** extracted sheaths containing the nanofiber network of **cable bacteria.** (A) EELS spectra of the carbon atoms at different pump laser powers. These spectra show three distinct peaks $C_1$, $C_2$ and $C_3$, attributed to $\pi^*, \sigma^*, SP^3$ hybridized peaks, respectively. The C1 peak becomes more pronounced at higher power, while C3 exhibits a blueshift of up to 3 eV as the power increases (shown by the dashed gray lines). (B) EELS spectra of nitrogen atoms, which has two $\pi^*$ and $\sigma^*$ peaks. Both peaks vanish at higher laser power. (C) EELS spectra of oxygen atoms show the $\sigma^*$ (C—O) peak, which fades and becomes redshifted at higher laser pump powers.

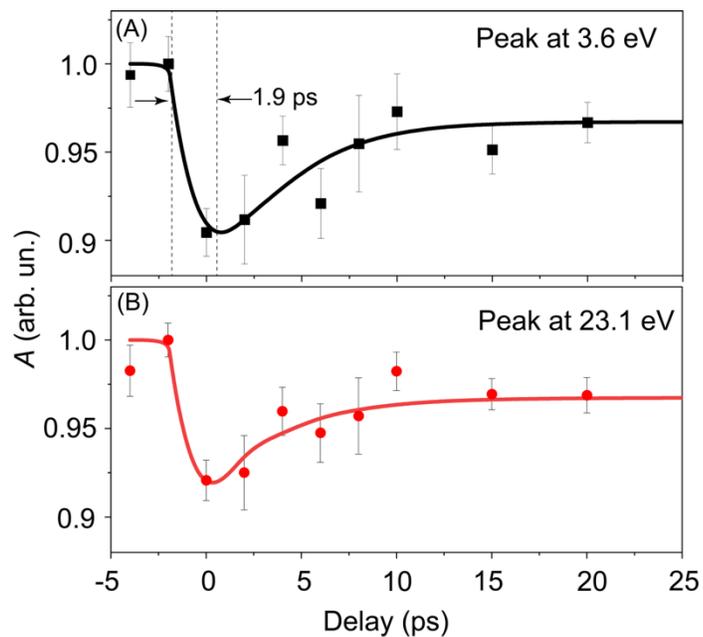

**Fig. 5. Ultrafast electron dynamics in the cable bacteria conductive network.** (**A and B**) The integration of the total number of electrons *A* of the π-electron (at 3.6 eV) and bulk plasmon-like (at 23.1 eV) peaks are shown in black and red points, respectively. The error bars represent the standard deviation of 5 scans. Biexponential fits of both dynamic curves are shown in black and red lines.